# Title: Terahertz-Light Driven Coupling of Antiferromagnetic Spins to Lattice

**Authors:** Evgeny A. Mashkovich[1,2]*, Kirill A. Grishunin[1,3], Roman M. Dubrovin[4], Anatoly K. Zvezdin[5,6], Roman V. Pisarev[4], and Alexey V. Kimel[1]

**Affiliations:**
[1]Radboud University, Institute for Molecules and Materials; Nijmegen 6525 AJ, The Netherlands
[2]University of Cologne, Institute of Physics II; Cologne D-50937, Germany
[3]MIREA - Russian Technological University; Moscow 119454, Russia
[4]Ioffe Institute, Russian Academy of Sciences; St Petersburg 194021, Russia
[5]Prokhorov General Physics Institute, Russian Academy of Sciences; Moscow 119991, Russia
[6]Moscow Institute of Physics & Technology; Dolgoprudnyi 141700, Russia

*Corresponding author. Email: mashkovich@ph2.uni-koeln.de

**Abstract:** Understanding spin-lattice coupling represents a key challenge in modern condensed matter physics, with crucial importance and implications for ultrafast and 2D-magnetism. The efficiency of angular momentum and energy transfer between spins and the lattice imposes fundamental speed limits on the ability to control spins in spintronics, magnonics and magnetic data storage. We report on an efficient nonlinear mechanism of spin-lattice coupling driven by THz light pulses. A nearly single-cycle THz pulse resonantly interacts with a coherent magnonic state in the antiferromagnet $CoF_2$ and excites the Raman-active THz phonon. The results reveal the unique functionality of antiferromagnets allowing ultrafast spin-lattice coupling using light.

**One-Sentence Summary:** Terahertz light pulses are used to coherently couple antiferromagnetic spins to the lattice.

**Main Text:**
Understanding how to control the spin-lattice interaction is a cornerstone for several hot topics of contemporary magnetic research, including ultrafast magnetism(*1*) and phononic control of magnetism(*2–6*), 2D magnetism(*7*), magnonics(*8–10*) and spintronics(*11*). Although this interaction is well understood in the vicinity of thermodynamic equilibrium, large-amplitude oscillations of atoms result in an essentially anharmonic lattice dynamics with an increasingly important role of higher order terms in the atomic free energy(*12*). For instance, driving coherent lattice vibrations into an anharmonic regime can promote energy transfer between different, otherwise non-interacting phononic modes via fully coherent phonon-phonon interactions(*12*). Similarly, one can expect a nonlinear mechanism of light-driven phonon-magnon coupling(*13–15*) and even anticipate non-trivial ultrafast phenomena associated with the physics of the Einstein–de Haas effect(*16*). Although several nonlinear mechanisms of phononic control of magnetism have been demonstrated theoretically(*3*) and experimentally(*4–6, 17*), alongside intense research interest devoted to exploring THz magnonics, understanding the nonlinear mechanism of energy transfer from THz magnons to THz phonons remains outstanding.

Antiferromagnets represent an appealing playground for the search of novel channels of spin-lattice coupling in the THz regime. Their spin structure can be modeled in the simplest case by two antiparallel sublattices with the net magnetizations $\mathbf{M}_1$ and $\mathbf{M}_2$, $|\mathbf{M}_1| = |\mathbf{M}_2|$. Alternatively, for describing the magnetic order, it is convenient to introduce the net magnetization vector $\mathbf{M} = \mathbf{M}_1 + \mathbf{M}_2$ and the antiferromagnetic (Néel) vector $\mathbf{L} = \mathbf{M}_1 - \mathbf{M}_2$. Typically, the frequencies of spin resonances in antiferromagnets are close to those of THz optical phonons. The objective of this study is therefore to use intense nearly single-cycle THz pulses(*18*) to drive coherent spin oscillations(*9*) and promote interactions between otherwise non-interacting magnonic and phononic modes.

To demonstrate light-driven spin-lattice coupling, we selected cobalt difluoride $CoF_2$ single crystal plate with the tetragonal rutile crystal structure. Below the Néel temperature $T_N = 39$ K, $CoF_2$ is a collinear antiferromagnet with a strong piezomagnetic effect(*19*). In an unit cell, the spins of $Co^{2+}$ ions at the cell's center are antiparallel to those at the cell's corners(*20*). These two types of ions form the two antiferromagnetic sublattices with the net magnetizations $\mathbf{M}_1$ and $\mathbf{M}_2$, respectively. The magnetizations are aligned along the crystallographic fourfold optical *z*-axis which is the "easy-axis" of magnetic anisotropy (see Fig. 1). The frequency of antiferromagnetic resonance is centered at $\Omega_m = 1.14$ THz at T = 6 K, while the nearest phonon mode of the $B_{1g}$ symmetry is at $\Omega_{ph} = 1.94$ THz(*21, 22*). Strong piezomagnetic properties of $CoF_2$ imply that atomic and spin dynamics can, in principle, be coupled(*4, 19*), but since the energies of the phonon and the magnon are substantially different, the coupling is inefficient. In this paper we reveal that a THz photon can fill the magnon-phonon energy gap and thus induce their efficient coupling. More specifically, we show that a nearly single-cycle THz pulse centered at ~1 THz with a bandwidth in excess of $2\Omega_m$-$\Omega_{ph}$ is able to prepare a coherent magnonic state and subsequently interact with this coherent state by promoting an energy transfer from the coherent magnon to the $B_{1g}$ phonon.

In order to trace this energy transfer, we employ a pump-probe technique to optically detect the coherent phonons and magnons. The atomic motion of the Raman active $B_{1g}$ phonon mode dynamically breaks the equivalence between the crystallographic *x*- and *y*- axes and thus induces linear birefringence for the light propagating along the optical *z*-axis. Moreover, upon introducing a generalized phonon coordinate $Q_{x^2-y^2}$ corresponding to the $B_{1g}$ phonon and the *z*-component of the Néel vector $L_z$, it can be shown that if the phonon $Q_{x^2-y^2}$ induces strain $\sigma_{xy}$, then the product $L_z\sigma_{xy}$ and the *z*-component of the magnetization $M_z$ transform equivalently under the symmetry operations allowed by the crystallographic 4/mmm point group of $CoF_2$ (The phonon selection rules)(*23*). This relation can also be attributed to the piezomagnetic effect(*19*), i.e. at temperatures below the Néel temperature, coherent atomic motion can also induce magnetic circular birefringence and result in the magneto-optical Faraday effect for light propagating along the *z*-axis. The polarization rotation must be proportional to $L_z\sigma_{xy}$ and therefore should disappear above the Néel temperature. The magnon mode at the frequency of the antiferromagnetic resonance can also be detected optically via magnetic circular birefringence, i.e. the Faraday effect, and also via magnetic linear birefringence (The magnon selection rules)(*23*).

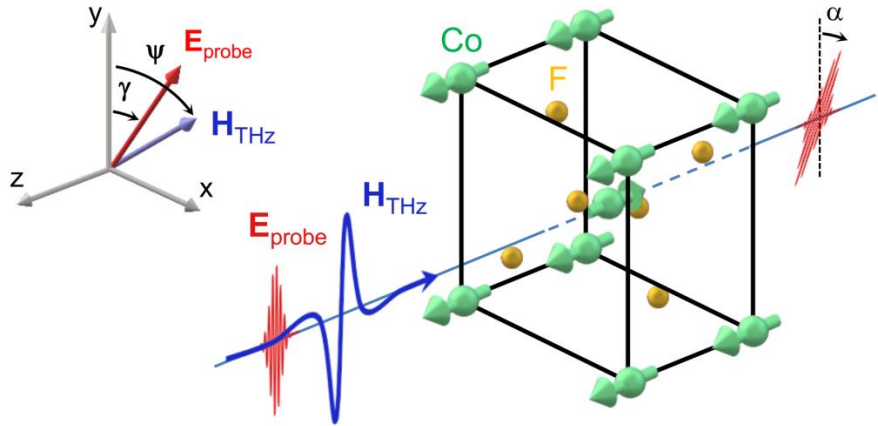

**Fig. 1. Geometry of the experiment and THz pumping of antiferromagnetic $CoF_2$.** The spins of $Co^{2+}$ ions are antiferromagnetically aligned along the *z*-axis. The linearly-polarized THz pump and near infrared probe beams propagate approximately collinearly along the *z*-axis and spatially overlap on the sample. By varying the time delay $t_{det}$ between the infrared probe and the THz pump pulses, we measure pump-induced ultrafast dynamics. The angle ψ between the THz pulse magnetic field $H_{THZ}$ and the *y*-axis is tuned in the range of ±π/2 using two wire-grid polarizers. The polarization of the probe pulse forms an angle γ with the *y*-axis and is controlled by a half-wave plate.

The typical THz-pump-induced transients for the probe's polarization rotation at different temperatures are shown in Fig. 2A. Polarization rotation with an amplitude of 150 mdeg is observed below the Néel point and reveals a strong temperature dependence. When normalized to the thickness of the crystal the signal reaches 3 deg/cm for a THz field on the order of MV/cm, which is comparable with the values reported for similar measurements on NiO(*24*). Figure 2B shows the Fourier transform of the time trace obtained at $T = 30$ K. It is seen that the THz pump pulse excites two resonances centered at 1.05 THz and 1.94 THz. Figure 2C shows that the temperature dependencies of the resonance frequencies in the vicinity of the Néel point are in a perfect agreement with the behaviors expected for the magnon and the $B_{1g}$ phonon in $CoF_2$(*25*,

26). While the magnon mode softens near the Néel point, the phonon frequency does not show any noticeable change in this temperature range. Figure 2D shows the dependencies of the amplitude of the magnon and the phonon resonances on the THz magnetic field strength. The magnon amplitude is a linear function of the field being typical for the conventional mechanism of excitation of the antiferromagnetic resonance by a magnetic field via the Zeeman torque(8, 27). In contrast, the phonon amplitude scales quadratically with the field strength similarly to Ref.(28), thus clearly evidencing the nonlinear mechanism of the excitation. Spectrogram of the probe polarization rotation at $T = 31$ K is plotted in fig. S10.

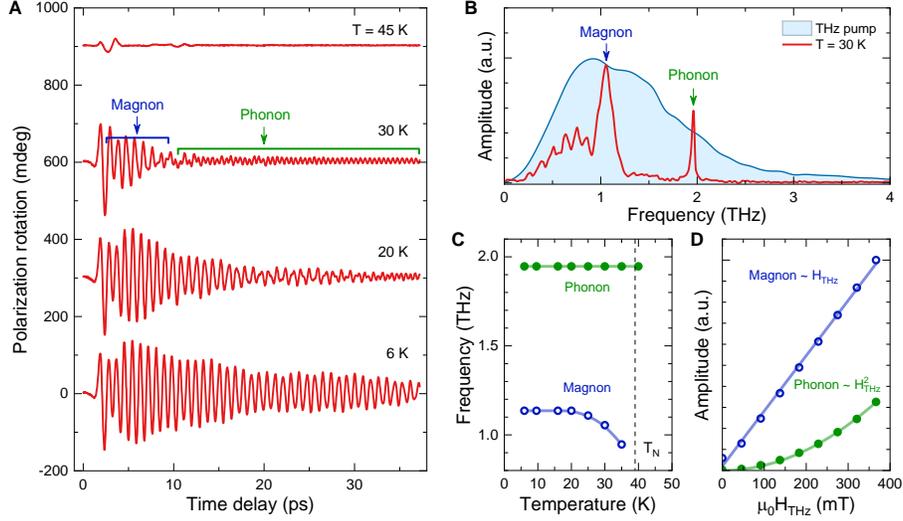

**Fig. 2. THz excitation of magnons and phonons.** (**A**) THz-pump-induced rotation of the probe's polarization measured at different temperatures for the case of the horizontal orientation of the probe electric field ($\gamma = 90°$) and the THz magnetic field ($\psi = 90°$). (**B**) Fourier spectrum of the waveform from panel (**A**) measured at $T = 30$ K. The spectrum of the incident THz pulse is shown by the light-blue area. (**C**) The frequencies of magnon (blue circles) and phonon (green dots) modes deduced from the Fourier spectrum as a function of temperature. The solid lines are guides for the eye. The dashed line marks the Néel temperature. (**D**) The Fourier spectral amplitudes of the magnon (blue circles) and phonon modes (green dots) at $T = 6$ K as functions of the THz magnetic field $H_{\text{THz}}$. The solid lines represent linear and quadratic fits. The corresponding waveforms are plotted in fig. S8 and S9.

In order to describe the dynamics of the Néel vector **L**, one can represent it as a sum of the stationary **L₀** and the THz induced **l** parts: **L** = **L₀** + **l**, where **L₀** >> **l**. From the Lagrange-Euler equations(29, 30) one finds that the dynamics of **l** is described by differential equations for damped harmonic oscillators (The magnon selection rules)(23). The latter gives

$$\begin{aligned} l_x(t) &\sim \tilde{H}(\Omega_m)\cos(\Omega_m t + \xi)e^{-t/\tau_m}\cos\psi, \\ l_y(t) &\sim \tilde{H}(\Omega_m)\cos(\Omega_m t + \xi)e^{-t/\tau_m}\sin\psi, \end{aligned} \quad (1)$$

where $\tilde{H}(\Omega_m)$ is the spectral amplitude of the THz magnetic field at $\Omega_m$, $\tau_m$ is the magnon damping time, and $\xi$ is the phase. Hence, any orientation of the magnetic field in the $xy$-plane excites oscillations of **l**. Note that the equations of motion for **l**, which can be derived either from the Lagrange-Euler or directly from the Landau-Lifshitz-Gilbert(*31*) equations, show that these oscillations are launched in the plane orthogonal to $\mathbf{H}_{THz}$.

In the case of atomic motion at the $B_{1g}$ phonon frequency with the generalized phononic coordinate $Q_{x^2-y^2}$, one can write a conventional equation of motion for the harmonic oscillator(*12*)

$$\frac{d^2 Q_{x^2-y^2}}{dt^2} + \frac{2}{\tau_{ph}} \frac{dQ_{x^2-y^2}}{dt} + \Omega_{ph}^2 Q_{x^2-y^2} = T_{x^2-y^2}, \quad (2)$$

where the first term in the left part corresponds to acceleration and the second term accounts for damping where $\tau_{ph}$ is the phonon damping time. The third term corresponds to the restoring force and $T_{x^2-y^2}(t)$ is the driving torque. Obviously, the equation must be invariant under all symmetry operations of the 4/mmm point group of CoF$_2$ i.e. the torque $T_{x^2-y^2}(t)$ must transform equivalently to $Q_{x^2-y^2}$ (see Table S1). Taking into account the irreducible representations for this point group(*32*) and focusing on terms of the second order with respect to the THz electric and magnetic fields, the following four contributions to $T_{x^2-y^2}(t)$ are allowed:

$$T_{x^2-y^2}(t) = C_1(E_x^2(t) - E_y^2(t)) + C_2(H_x^2(t) - H_y^2(t)) + C_3(l_x(t)H_y(t) - l_y(t)H_x(t)) + C_4(l_x^2(t) - l_y^2(t)), \quad (3)$$

where $C_1$, $C_2$, $C_3$, and $C_4$ are phenomenological coefficients. The first and the second terms correspond to electric and magnetic dipole mechanisms of two-photon excitation of the $B_{1g}$ phonon via a virtual state. A similar mechanism was described in Refs.(*33, 34*). In contrast to the two-photon excitation of phonons via intermediate phononic state, as proposed in Ref. (*35*), the third term in Eq.3 describes two-photon excitation of phonons via an intermediate magnonic state. The fourth term shows that the phonon can also be excited because of the anharmonicity of magnons. However, one may argue that in the case of forced oscillations of the antiferromagnetic Néel vector by a THz magnetic field, the effects of the third and the fourth terms must be similar and can hardly be distinguished.

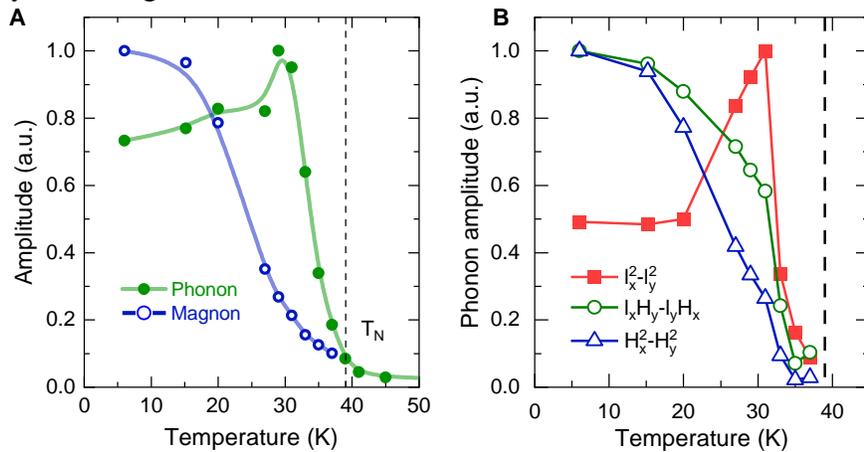

**Fig. 3. Phonon and magnon excitation.** (**A**) Temperature dependencies of the Fourier amplitudes of the magnon (blue circles) and the phonon (green dots) modes. The solid lines are guides to the eye. The dashed line marks the Néel temperature. The corresponding waveforms are given in Supplemental material in fig. S8. (**B**) Estimated phononic amplitude excited via the virtual state (blue triangles; first and second terms in Eq.3), real magnonic state (green circles; the third term in Eq. 3) and anharmonic magnon (red squeres; the fourth term in Eq.3).

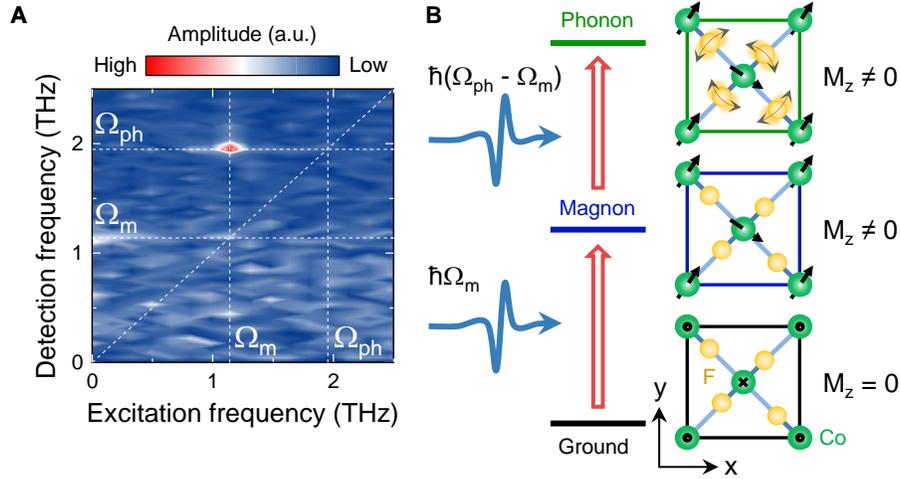

**Fig. 4**. **THz light-driven coupling of lattice to spins**. (**A**) Two-dimensional (2D) Fourier spectrum of the nonlinear amplitude $\tilde{\alpha}_{NL}(f_{exc}, f_{det})$. The measurements were performed at $T = 10$ K for horizontal orientations of the probe electric field ($\gamma = 90°$) and the THz magnetic field ($\psi = 90°$). (**B**) A pictorial of the magnon-mediated excitation of the $B_{1g}$ phonon by a THz magnetic field. The unperturbed antiferromagnetic state of the $CoF_2$ unit cell is shown at the bottom. A THz photon resonantly populates the coherent magnonic state at the frequency $\Omega_m$ thus creating an intermediate state. Another THz photon at the frequency of $\Omega_{ph} - \Omega_m$ interacts with this intermediate state and coherently excites the $B_{1g}$ phonon.

The temperature dependencies of the detected signal at the phonon and magnon frequencies are shown in Fig. 3A. The substantially low (~ 30 times) phononic oscillations above the Néel point is a result of the dominating role of the magneto-optical Faraday effect in the detection of the phonon. Intriguingly, the detected phononic amplitude stays nearly constant in the range from 6 to 25 K and even has the maximum at 30 K, where the equality $\Omega_{ph} = 2\Omega_m$ holds. In contrast, the magnon frequency $\Omega_m$ smoothly and monotonically decreases from 1.14 THz at $T = 6$ K to 0 at the Néel point, and this frequency change is accompanied by broadening of the magnon spectral line $\Delta\Omega_m$ (see fig. S3). Such changes affect differently the efficiencies of phononic excitations due to the mechanisms corresponding to different terms in Eq. 3. To elucidate this, we modelled temperature dependencies of the mechanisms corresponding to the first, third and the fourth term in Eq. 3. In particular, our model explored how temperature affects the strength of the oscillations induced at the frequency of the phonon (The phonon selection rules)(*23*). In the modelling, we also took into account that the phonon and magnon are detected optically and the sensitivity of the detection is practically proportional to the antiferromagnetic Néel vector. Comparing the simulated dependencies with that observed experimentally clearly emphasizes the dominating role of the magnon-mediated mechanism in the excitation of the phonon (see Fig. 3B).

In order to demonstrate that the non-linear excitation of phonons is mediated by coherent magnonic state, we performed double-pump experiments using two identical THz pulses separated by a time delay $t_{exc}$ with a peak magnetic field of 100 mT. For different $t_{exc}$, we measured the probe's polarization rotation as a function of time delay between the first THz pump and the probe pulse $t_{det}$. Employing the same method of data acquisition as explained in Ref.(*36*), we measured the dynamics triggered by both THz-pump pulses $\alpha_{oo}(\tau)$, by the first pump only, i.e. when the second THz pump is blocked $\alpha_{oc}(\tau)$ and by the second THz pump only (the first THz pump is blocked) $\alpha_{co}(\tau)$. Here $\tau$ is the vector $\tau = (t_{exc}; t_{det})$. We deduce the intrinsically nonlinear response of the medium by subsequently calculating the difference between the signals in the time-domain:

$$\alpha_{NL}(\tau) = \alpha_{oo}(\tau) - \alpha_{oc}(\tau) - \alpha_{co}(\tau), \tag{4}$$

where $\alpha_{NL}(\tau)$ is non-zero only if the first pump changes the spin susceptibility to the THz magnetic field. A Fourier transform of the extracted signals $\alpha_{NL}(\tau)$ allows us to plot the spectrum as a 2D graph $\tilde{\alpha}_{NL}(f_{exc}; f_{det})$ (see Fig. 4A). Here $f_{det}$ and $f_{exc}$ are the Fourier frequencies of the corresponding time delays $t_{det}$ and $t_{exc}$, respectively. The spectrum $\tilde{\alpha}_{NL}(f_{exc}; f_{det})$ clearly reveals one maximum at $f_{det} = \Omega_{ph}$ and $f_{exc} = \Omega_m$, revealing that the excitation of the $B_{1g}$ phonon via a nonlinear terahertz-light-driven mechanism is only possible when the magnon is excited. This means that the macroscopic coherent magnonic state with the frequency $\Omega_m$, resonantly generated by the broadband THz pulse, has an essentially different susceptibility to the THz magnetic field compared to that associated with its unperturbed state. More particularly, the interaction of the second THz photon at the frequency $\Omega_{ph} - \Omega_m$ with the coherent magnonic state is able to excite the $B_{1g}$ phonon at the frequency $\Omega_{ph}$ (see Fig. 4B).

The ultrafast coherent transfer of spin energy to the lattice driven by THz light pulses opens up possibilities for the fields of nonlinear phononics, ultrafast magnetism, THz magnonics and antiferromagnetic spintronics. The dynamic coupling of the lattice to the spins is crucial for ultrafast control of magnetism and phase engineering through transient changes of magnetic and structural states. Exactly the same light-driven spin-lattice coupling is allowed in other antiferromagnetic fluorides having the same point group ($MnF_2$, $FeF_2$ et al.), but the effect must be more general and can be extended to other materials. As a coherent magnonic state in any ferro-, ferri- and even antiferromagnetic materials can induce dynamic magnetization, the excitation of this coherent state with a magnetic field at the frequency matching the gap between the magnon and the phonon might even lead to non-trivial ultrafast phenomena associated with the physics of the Einstein-de Haas effect. This mechanism is especially appealing for ultrafast coherent control of materials with complex charge and spin ordering, for example, in multiferroics and 2D magnets.

**Acknowledgments:** The authors thank Dr. Sergey Semin and Chris Berkhout for technical support. The authors acknowledge fruitful discussion with Dr. Mikhail Prosnikov and Dr. Carl Davies.

**Funding:**

The research was funded by the Netherlands Organization for Scientific Research (NWO).

The contribution of E.A.M has been funded by the Deutsche Forschungsgemeinschaft (DFG, German Research Foundation) - Project number 277146847 - CRC 1238.

K.A.G. acknowledges support from Russian Foundation for Basic Research (grant 18-02-40027).

R.M.D. and R.V.P. acknowledge the financial support by Russian Foundation for Basic Research according to the project No. 19-02-00457.

The theoretical contribution of A.K.Z. has been supported by Russian Science Foundation (the Project N 17-12-01333).

**Author contributions:**

Conceptualization: AEM, RVP, AVK

Methodology: AEM, KAG, AKZ

Investigation: AEM, KAG, RMD

Resources: RMD, RVP

Supervision: AVK

Writing – original draft: AEM

Writing – review & editing: AEM, KAG, RMD, AKZ, RVP, AVK

**Competing interests:** Authors declare that they have no competing interests.

**Data and materials availability:** All data are deposited at Zenodo(47).

**Supplementary Materials**

Materials and Methods

Supplementary Text

Fig S1 to S10

Table S1

References (37–46)

**Materials and Methods**

Terahertz pump-probe spectroscopy

In our experiment, a linearly polarized THz pump was generated by the tilted-front optical rectification in a LiNbO$_3$ prism of an amplified 100 fs laser pulse at the central wavelength of 800 nm (*18*, *37*). Electro-optical sampling in a 50 μm thick [110]-cut GaP crystal was used to map the electric field of the generated THz pulses. The peak THz electric field strength was up to 1100 kV/cm and the corresponding magnetic induction field was up to 360 mT. The THz spectrum of the pulse was centered at 0.9 THz with the full width at half maximum of 1.5 THz (see fig. S1). To detect the THz-pulse induced dynamics in CoF$_2$ sample, we used a linearly polarized (100 fs, 800 nm) probe laser pulse and measured its polarization rotation experienced upon propagation through the studied sample. The lateral size of the THz beam spot is considerably larger than the optical one. The experiments were performed in a dry ambient atmosphere to remove water absorption lines in the THz spectrum. Orientation of THz pulse polarization and electric field strength were controlled by two wire-grids polarizes. The CoF$_2$ sample was cut as 507±10 μm-thick plate from a single crystal with the *z*-axis at ~5° with respect to the sample normal. The pump and probe beams were directed along the sample normal.

Two-dimensional terahertz spectroscopy

In order to upgrade the pump-probe setup to the one for 2D THz spectroscopy, we split the pump pulse into two beams and control the delay $t_{exc}$ between the pulses with an additional mechanical delay line. Afterwards the pump pulses are brought together and aligned in the one pump beam again. THz generation and characterization were done as explained for the single THz pump-probe spectroscopy. The maximum THz electric field strength was up to 300 kV/cm, which corresponds to its 100 mT magnetic field counterpart. For each fixed delay $t_{exc}$, the THz induced probe polarization rotation $\alpha(t_{exc}, t_{det})$ was measured using balanced detection scheme. Both the two THz pumps and the probe beams were directed along the sample normal.

Choosing different $t_{exc}$, we detected polarization rotation as a function of the time delay between the first THz pump and the probe pulse $t_{det}$. We used different repetition rates for the probe pulse (1 kHz), the first THz pump pulse (500 Hz) and the second THz pump pulse (250 Hz). Employing data acquisition card as explained in (36), we measured the dynamics triggered by the both THz-pump pulses $\alpha_{oo}$, by the first THz pump only i.e. when the second THz pump is chopped away $\alpha_{oc}$, by the second THz pump only (the first THz pump is chopped away) $\alpha_{co}$ and the noise $\alpha_{cc}$. According to the algorithm explained in (38) we calculated the dynamics triggered by the both THz-pump pulses with subtracted noise

$$\alpha_2 = \alpha_{oo} - \alpha_{cc} \qquad (1)$$

and the nonlinear part of THz induced dynamics

$$\alpha_{NL} = (\alpha_{oo} - \alpha_{cc}) - (\alpha_{oc} - \alpha_{cc}) - (\alpha_{co} - \alpha_{cc}). \qquad (2)$$

Performing 2D Fourier transform on $\alpha_2(t_{exc}, t_{det})$ and $\alpha_{NL}(t_{exc}, t_{det})$, we get 2D Fourier spectra $\tilde{\alpha}_2(f_{exc}, f_{det})$ and $\tilde{\alpha}_{NL}(f_{exc}, f_{det})$, correspondingly. Here $f_{det}$ and $f_{exc}$ are Fourier frequencies of the corresponding time delays $t_{det}$ and $t_{exc}$. To exclude mutual influence of the two optical pumps on THz generation done in the same LiNbO$_3$ crystal, $t_{exc}$ variations were studied after 2 ps. Initial position of $t_{det}$ was stiffed together with the $t_{exc}$ change to focus on the effect of the second THz pump on dynamics induced by the first THz pump.

Figure S2 shows the 2D Fourier spectrum $\tilde{\alpha}_2(f_{exc}, f_{det})$. The peaks along the $f_{exc}=0$ line correspond to the single THz pump-probe spectrum. Except these peaks the strongest maxima observed in the $\tilde{\alpha}_2(f_{exc}, f_{det})$ can be attributed to linear excitation of the magnon ($f_{exc}=\Omega_m$ and $f_{det}=\Omega_m$), nonlinear magnon-mediated excitation of the phonon ($f_{exc}=\Omega_m$ and $f_{det}=\Omega_{ph}$), and interference of the phonon modes excited by the first and the second THz pump pulses ($f_{exc}=\Omega_{ph}$ and $f_{det}=\Omega_{ph}$).

**The magnon mode selection rules**

<u>Ultrafast spin dynamics and Lagrangian formalism</u>

Following the theoretical formalism of Ref.(39), equations describing magnetization dynamics of a collinear antiferromagnet can be derived from the Lagrangian $\Lambda$, the dissipative Reyleigh function $R$, and Lagrange-Euler equations

$$\Lambda = \frac{\chi_\perp}{2}\left[\left(\frac{\dot\theta}{\gamma}+H_x\sin\varphi\right)^2 + \left(\left(\frac{\dot\varphi}{\gamma}-H_z\right)\sin\theta + H_x\cos\theta\cos\varphi\right)^2\right] - W_{ef}(\mathbf{s}), \quad (3)$$

$$R = \frac{\alpha M_0}{\gamma}\left(\dot\theta^2 + \sin^2\theta\dot\varphi^2\right), \quad (4)$$

$$\frac{d}{dt}\frac{\partial\Lambda}{\partial\dot q} - \frac{\partial\Lambda}{\partial q} = -\frac{\partial R}{\partial\dot q} \quad (q=\theta \text{ or } \varphi), \quad (5)$$

where $\varphi$ and $\theta$ angles determine orientation of the normalized Néel vector $\mathbf{s}=\mathbf{L}/(2M_0)$,

$$\mathbf{s}=(\sin\theta\cos\varphi \quad \sin\theta\sin\varphi \quad \cos\theta). \quad (6)$$

In-plane components of the THz pulse magnetic field are $H_x=H_{THz}\sin\psi$ and $H_z=H_{THz}\cos\psi$. The magnetic free energy is

$$W = \frac{\varepsilon m^2}{2} + as_y^2 - d(m_x s_z + m_z s_x), \quad (7)$$

where $\varepsilon$, $a$ and $d$ denote exchange, anisotropy and Dzyaloshinskii constants, correspondingly. $\mathbf{m}=\mathbf{M}/(2M_0)$ is the normalized net magnetization. To resolve degeneracy of the excited magnon mode we use coordinate system with *xz*-sample plane and out-of-plane *y*-axis. It will be shown further that in this case dynamics of $\varphi$ and $\theta$ can be separately attributed to the certain eigen frequencies. Minimization of Eq. 7 with respect to $m_x$ and $m_z$ results in

$$m_x = \frac{d}{\varepsilon}s_z \text{ and } m_z = \frac{d}{\varepsilon}s_x. \quad (8)$$

Substituting Eq. 8 in Eq. 7 gives

$$W_{ef}(\mathbf{s}) = \tilde{a}s_y^2 = \tilde{a}\sin^2\theta\sin^2\varphi. \tag{9}$$

Minimizing the free energy $W_{ef}(\mathbf{s})$ on $\theta$ and $\varphi$ determines the ground states: $\theta_0 = \pi/2$ and $\varphi_0 = \pi/2$ (or $3\pi/2$). After substitution of Eqs. 3-4 into Eq. 5 and linearization near the ground state we get

$$\frac{d^2\theta_1}{dt^2} + \frac{2}{\tau_m}\frac{d\theta_1}{dt} + \Omega_m^2\theta_1 = -\gamma\frac{dH_x}{dt} \text{ and } \frac{d^2\varphi_1}{dt^2} + \frac{2}{\tau_m}\frac{d\varphi_1}{dt} + \Omega_m^2\varphi_1 = \gamma\frac{dH_z}{dt}, \tag{10}$$

where perturbations near the ground state are $\theta_1 = \theta - \theta_0$ and $\varphi_1 = \varphi - \varphi_0$. The frequency of antiferromagnetic resonance is $\Omega_m = \sqrt{2|\tilde{a}|\gamma^2/\chi_\perp}$. From here it is seen that the THz pump excites(40,41) doubly degenerat antiferromagnetic mode. It is worth noting, that $\Omega_m$ is in perfect agreement with earlier theoretical representation if we take into account that the anisotropy field $H_A = 2\tilde{a}/M_0$ and the exchange field $H_E = M_0/\chi_\perp$(42,43). Also note that $\varphi_0$ determines 180-degree domain orientation, which size ~ 100 μm(44) is considerably larger than the lateral size of the optical probe beam. Summing up oscillations of $\varphi_1$ and $\theta_1$ result in the THz induced perturbation of the Néel vector:

$$l'_x = 2\gamma M_0 H(\Omega_m)\cos(\Omega_m t + \zeta)\exp(-t/\tau_m)\cos\psi,$$
$$l'_z = -2\gamma M_0 H(\Omega_m)\cos(\Omega_m t + \zeta)\exp(-t/\tau_m)\sin\psi \tag{11}$$

near the stationary ground state

$$L_z = L_{0z} = 2M_0, \tag{12}$$

where $\tan\zeta = 1/(\Omega_m\tau_m)$ and $H(\Omega_m)$ is spectral amplitude of the THz pulse at the antiferromagnetic resonance frequency $\Omega_m$. Note, that in the laboratory coordinate system, the normal to the sample corresponds to the z-axis and the xy-plane coincides with the sample plane i.e. $l'_x \to l_x$ and $l'_z \to -l_y$.

Dielectric permittivity modulation by the magnon mode.

Probe polarization rotation induced by the magnon mode was analyzed using the principles of thermodynamics and symmetry arguments. The fourth rank polar i-tensor $\chi'_{ijkl}$ and the third rank axial i-tensor $\chi'_{ijk}$ determine Faraday and Cotton-Mouton effects, respectively. In crystallographic coordinate system $x'y'z'$ for 4/mmm point group of $CoF_2$ crystal the tensors have the following forms(45)

$$\chi'_{ijkl} = \begin{pmatrix} C_{11} & C_{21} & C_{13} & 0 & 0 & 0 \\ C_{21} & C_{11} & C_{13} & 0 & 0 & 0 \\ C_{31} & C_{31} & C_{33} & 0 & 0 & 0 \\ 0 & 0 & 0 & C_{55} & 0 & 0 \\ 0 & 0 & 0 & 0 & C_{55} & 0 \\ 0 & 0 & 0 & 0 & 0 & C_{66} \end{pmatrix}, \tag{13}$$

where *ij*-indexes are substituted by $i$ and $9-(i+j)$ if $i=j$ and $i \neq j$, respectively. The same procedure is applied for *kl*-indexes.

$$\chi'_{ijk} = \begin{pmatrix} 0 & 0 & 0 \\ 0 & 0 & F_{123} \\ 0 & F_{132} & 0 \\ 0 & 0 & -F_{123} \\ 0 & 0 & 0 \\ -F_{132} & 0 & 0 \\ 0 & -F_{132} & 0 \\ F_{132} & 0 & 0 \\ 0 & 0 & 0 \end{pmatrix}, \tag{14}$$

where the column indexing are 11, 12, 13, 21, 22, 23, 31, 32 and 33.

The laboratory coordinate system is depicted in Fig. 1 of the main text, where the *xy*-plane is a sample surface and the *z*-axis is a surface normal. The studied $CoF_2$ plate was cut from a single crystal such that the $z'$-axis was at about 5° to the normal of sample. Transformation from the crystallographic to the laboratory coordinate system goes via rotation around $x'$- and $y'$-axis by the two small angles $\theta_x$ and $\theta_y$, correspondingly, and is determined by the rotational matrix $R_{xy} = R_x R_y$, where $R_x$ and $R_y$ are

$$R_x = \begin{pmatrix} 1 & 0 & 0 \\ 0 & 1 & -\theta_x \\ 0 & \theta_x & 1 \end{pmatrix}, \tag{15}$$

$$R_y = \begin{pmatrix} 1 & 0 & \theta_y \\ 0 & 1 & 0 \\ -\theta_y & 0 & 1 \end{pmatrix}. \tag{16}$$

After transformation to the laboratory coordinate system the main magnetic contribution in the symmetric part of dielectric permittivity $\epsilon^{(s)}_{ij}$ is determined by

$$\begin{aligned}\epsilon^{(s)}_{11} &= \chi_{11kl} L_{0k} l_l = L_{0z} l_x (\chi_{1113} + \chi_{1131}) + L_{0z} l_y (\chi_{1123} + \chi_{1132}), \\ \epsilon^{(s)}_{22} &= \chi_{22kl} L_{0k} l_l = L_{0z} l_x (\chi_{2213} + \chi_{2231}) + L_{0z} l_y (\chi_{2223} + \chi_{2232}),\end{aligned} \tag{17}$$

where

$$\chi(i,j,k,l) = R_{xy}(i,i') R_{xy}(j,j') R_{xy}(k,k') R_{xy}(l,l') \chi'(i',j',k',l') \tag{18}$$

with

$$\begin{aligned}\chi_{1113} + \chi_{1131} &= \theta_y (2C_{13} - 2C_{11} + 3C_{55}), \\ \chi_{1123} + \chi_{1132} &= \theta_x (2C_{21} - 2C_{13}), \\ \chi_{2213} + \chi_{2231} &= \theta_y (2C_{13} - 2C_{21}), \\ \chi_{2223} + \chi_{2232} &= \theta_x (2C_{11} - 2C_{13} - 3C_{55}).\end{aligned} \tag{19}$$

Note, that $L_{0z}$ and $l_{x,y}$ are the stationary and the THz induced parts of the Néel vector, respectively. After substitution of Eq. 19 into Eq. 17 we eventually can write

$$\epsilon_{11}^{(s)} = L_{0z}[Al_x\theta_y + Bl_y\theta_x],$$
$$\epsilon_{22}^{(s)} = -L_{0z}[Bl_x\theta_y + Al_y\theta_x]. \qquad (20)$$

Repeating the laboratory-crystallographic transformation procedure, anti-symmetric part of dielectric permittivity $\epsilon_{ij}^{(a)}$ can be presented as

$$\epsilon_{12}^{(a)} = \chi_{12k}M_k = M_x\chi_{121} + M_y\chi_{122}, \qquad (21)$$

where

$$\chi(i,j,k) = det(R_{xy})R_{xy}(i,i')R_{xy}(j,j')R_{xy}(k,k')\chi'(i',j',k') \qquad (22)$$

with

$$\chi_{121} = \theta_y(F_{123} + F_{132}),$$
$$\chi_{122} = -\theta_x(F_{123} + F_{132}). \qquad (23)$$

After substitution of Eq. 23 into Eq. 21

$$\epsilon_{12}^{(a)} = D[M_x\theta_y - M_y\theta_x] \qquad (24)$$

and taking into account that THz pulse induces the net magnetization $\mathbf{M} = -\frac{1}{\gamma H_E}[\mathbf{L} \times \frac{d\mathbf{L}}{dt}]$ according to Ref. (29,30)

$$M_x \sim -L_{0z}\dot{l}_y,$$
$$M_y \sim L_{0z}\dot{l}_x \qquad (25)$$

we can rewrite

$$\epsilon_{12}^{(a)} = DL_{0z}[\dot{l}_y\theta_y + \dot{l}_x\theta_x]. \qquad (26)$$

As a prove that magnetic contributions to permittivity indeed plays a role we measured static relative change of the probe polarization rotation after propagating through the sample as a function of temperature (see fig. S3).

Combining Eqs. 20, 26 and 11, modulations of symmetric and anti-symmetric permittivities can be written as

$$\epsilon_{11}^{(s)} = L_{0z}H(\Omega_m)[A\theta_y \sin\psi \cos\Omega_m t + B\theta_x \cos\psi \cos\Omega_m t],$$
$$\epsilon_{22}^{(s)} = L_{0z}H(\Omega_m)[B\theta_y \sin\psi \cos\Omega_m t + A\theta_x \cos\psi \cos\Omega_m t], \qquad (27)$$

and

$$\epsilon_{12}^{(a)} = DL_{0z}H(\Omega_m)[\theta_y \cos\psi \sin\Omega_m t + \theta_x \sin\psi \sin\Omega_m t]. \qquad (28)$$

Modulation of $\epsilon^{(s)}$ and $\epsilon^{(a)}$ results in the probe pulse polarization rotation $\alpha$, which is detected by a balanced detector. Assuming for simplicity independent contributions to the polarization rotation we can write(8)

$$\alpha = (J_1 \cos\psi + J_2 \sin\psi)\sin\Omega_m t + (J_3 \cos\psi + J_4 \sin\psi)\sin(4\gamma + 4\gamma_0)\cos\Omega_m t, \qquad (29)$$

where $\gamma$ is the probe pulse electric field orientation angle, $\gamma_0$ is the phase mismatch and $J_1$, $J_2$, $J_3$, $J_4$ are constants.

To define dependence of the probe polarization rotation $\alpha$ on orientations of the probe pulse electric field $\gamma$ and the THz magnetic field $\psi$ we introduce $\alpha_{probe}$ and $\alpha_{pump}$, which are determined as

$$\alpha_{pump} = \Pi_{pump} \cos(\Omega_m t - \beta_{pump}), \quad (30)$$

where

$$\Pi_{pump} = \sqrt{R_1^2 \sin^2(\psi + \psi_1) + R_2^2 \cos^2(\psi + \psi_2)},$$
$$\tan \beta_{pump} = \frac{R_1}{R_2} \frac{\sin(\psi + \psi_1)}{\cos(\psi + \psi_2)} \quad (31)$$

and

$$\alpha_{probe} = \Pi_{probe} \cos(\Omega_m t - \beta_{probe}), \quad (32)$$

where

$$\Pi_{probe} = \sqrt{R_1^2 + R_2^2 \sin^2(4\gamma + 4\gamma_0)},$$
$$\tan \beta = \frac{R_1}{R_2 \sin(4\gamma + 4\gamma_0)}. \quad (33)$$

Summing up, if the probe beam propagates at non-zero angle with respect to the $z$-axis, the pump-induced magnetization changes can be detected not only via magnetic circular birefringence, i.e. the Faraday effect, but also via magnetic linear birefringence.

**The phonon mode selection rules**

Symmetry formalism of phonon excitation

Symmetry implies that the torque, which triggers the $B_{1g}$ phonon mode, has the form

$$T_{x^2-y^2} = C_1(E_x^2(t) - E_y^2(t)) + C_2(H_x^2(t) - H_y^2(t)) + C_3(l_x(t)H_y(t) - l_y(t)H_x(t)) + C_4(l_x^2(t) - l_y^2(t)). \quad (34)$$

where $C_1$, $C_2$, $C_3$ and $C_4$ are phenomenological coefficients. The first two terms correspond to the two-photon electric and magnetic dipole excitation of the phonon via a virtual state similar to those described in Ref. (33,34). The last two terms render two photon magnon-mediated pumping of the phonon. The corresponding generalized phonon coordinate $Q_{x^2-y^2}$ can be determined from the damped harmonic oscillator equation(12)

$$\frac{d^2 Q_{x^2-y^2}}{dt^2} + \frac{dQ_{x^2-y^2}}{dt} \frac{2}{\tau_{ph}} + \Omega_{ph}^2 Q_{x^2-y^2} = T_{x^2-y^2}, \quad (35)$$

Different terms in Eq.34 contribute differently to the torque at the phonon frequency $\Omega_{ph}$, because they have different dependencies on the frequency:

$$T_{x^2-y^2}(\Omega_{ph}) \sim A\int H_{THz}(f) H_{THz}(f - \Omega_{ph}) df + B\int l(f) H_{THz}(f - \Omega_{ph}) df + C\int l(f) l(f - \Omega_{ph}) df, \quad (36)$$

where *A, B* and *C* are phenomenological coefficients. The first term is related to the first and the second terms in Eq. 34 and corresponds to excitation via virtual state; the second term is related to the third term in Eq. 34 and corresponds to pumping via magnonic state; the third term corresponds to the fourth one in Eq. 34 and originates from the magnon anharmonicity. The experimentally osereved tempertaure dependencies of the width *w* and frequency $f_0$ of the magnon line enter Eq. 36 via the Lorentzian

$$l(f) \sim \frac{w}{4(f-f_0)^2 + w^2}, \tag{37}$$

Temperature dependencies of $f_0(T)$ and $w(T)$ are plotted on Fig. 2C and fig. S4, correspondingly. In the modelling, we also took into account that the phonon and magnon modes are detected optically and the sensitivity of the detection is in fact proportional to the antiferromagnetic Néel vector. According to Eq. 29, the amplitude of the probe polarization rotation induced by the magnon mode

$$\alpha_{mag} \sim IL_{0z} \tag{38}$$

and the phonon mode

$$\alpha_{ph} \sim AL_{0z}(H_{THz}*H_{THz}) + BL_{0z}(I*H_{THz}) + CL_{0z}(I*I), \tag{39}$$

where * stays for convolution. The latter expression can also be presented differently substituting l by $\alpha_{mag}$ i.e. combining Eqs. 38 and 39

$$\alpha_{ph} \sim AL_{0z}(H_{THz}*H_{THz}) + B(\alpha_{mag}*H_{THz}) + C(\alpha_{mag}*\alpha_{mag})\frac{1}{L_{0z}} \tag{40}$$

The result of simulation of $\alpha_{ph}$ as a function of temperature is plotted in Fig. 3B.

Note that all four terms in Eq. 34 do depend on the polarization of the THz pulse and the dependence is decribed by harmonic function of $2\psi$. In order to account for polarization independent part, one should take into account that the length of the antiferromagnetic Néel vector is not conserved.

Dielectric permittivity modulation by the phonon mode

The group operators of 4/mmm point group are $4\underline{z}$ and $2d$ (32). The transformations under these operators are summarized in Table S1. If the phonon $Q_{x^2-y^2}$ induces strain $\sigma_{xy}$, it can be seen that $M_z$-projection of magnetization transforms in a similar way as $\sigma_{xy}L_{0z}$ under all symmetry operations of 4/mmm group. Hence, the modulation of anti-symmetric part of dielectric permittivity tensor can be written as

$$\epsilon^{(a)}_{12} = \chi_{123}M_z = F_{123}M_z = \sigma_{xy}L_{0z}F_{123}. \tag{41}$$

The modulation of symmetric part of dielectric permittivity modulation is determined from Eq. 13 as

$$\begin{aligned}\epsilon^{(s)}_{11} &= (\chi_{1111} - \chi_{1122})Q_{x^2-y^2} = (C_{11} - C_{21})Q_{x^2-y^2} \\ \epsilon^{(s)}_{22} &= (\chi_{2211} - \chi_{2222})Q_{x^2-y^2} = (C_{21} - C_{11})Q_{x^2-y^2}\end{aligned} \tag{42}$$

In the case of independent contribution of $\epsilon^{(s)}$ and $\epsilon^{(a)}$ to the probe polarization rotation we have:

$$\alpha = [Y_1 L_{0z} + Y_2 \sin(4\gamma + 4\gamma_0)]Q_{x^2-y^2}, \tag{43}$$

where $Y_1$ and $Y_2$ are constants. Let's consider for siplicity that THz pump only contains two spectrall components $\mathbf{H}_{THz}\cos\Omega_m t$ and $\mathbf{H}_{THz}\cos(\Omega_{ph} - \Omega_m)t$. Taking into account polarization independant contribution in the most general form $\sim \cos(\Omega_{ph}t + \varphi_0)$, where $\varphi_0$ is the phase, we get

$$\alpha = [Y_1 L_{0z} + Y_2 \sin(4\gamma + 4\gamma_0)][Y_3 \cos 2\psi \cos\Omega_{ph}t + Y_4 \cos(\Omega_{ph}t + \varphi_0)], \tag{44}$$

From here the phonon mode amplitude dependencies $\alpha_{probe}$ and $\alpha_{pump}$ on the orientation angles of the probe electric field $\gamma$ and the THz magnetic field $\psi$ can be expressed.

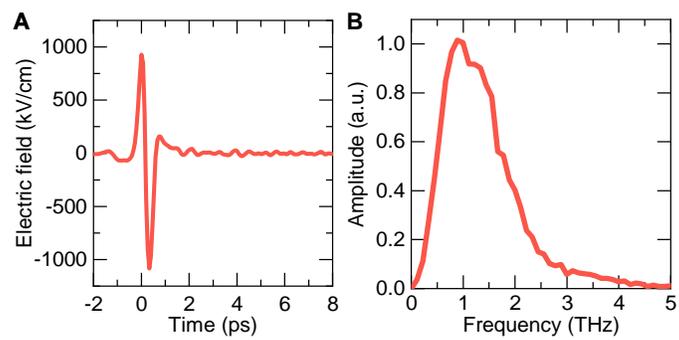

**Fig. S1.** (**A**) Terahertz pulse waveform. (**B**) Corresponding normalized Fourier spetrum.

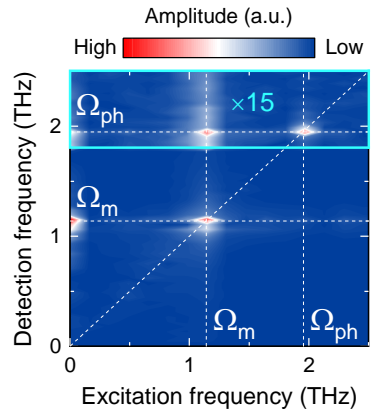

**Fig. S2.** Two-dimensional (2D) Fourier spectrum $\alpha_2(f_{exc}, f_{det})$.

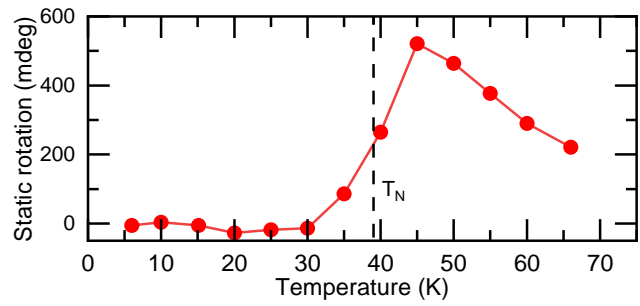

**Fig. S3.** Relative change of probe polarization rotation after propagating through the sample as a function of temperature. Electric field of probe pulse is horizontal γ = 90°. THz pump pulse is blocked. Change of the slope sign near $T_N$= 39 K indicates the presence of static magnetic birefringence in the sample(45).

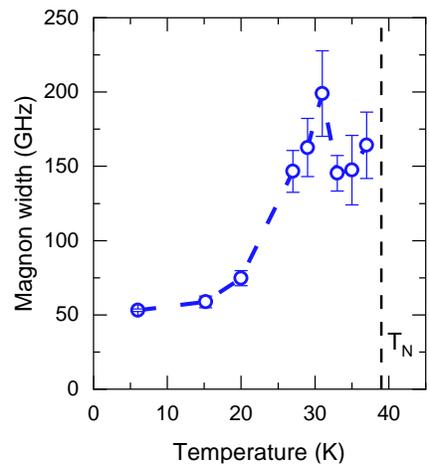

**Fig. 4.** The magnon spectral line width as a function of temperature. The width is estimated from data shown in fig. S8B fitting the data with Lorentzian.

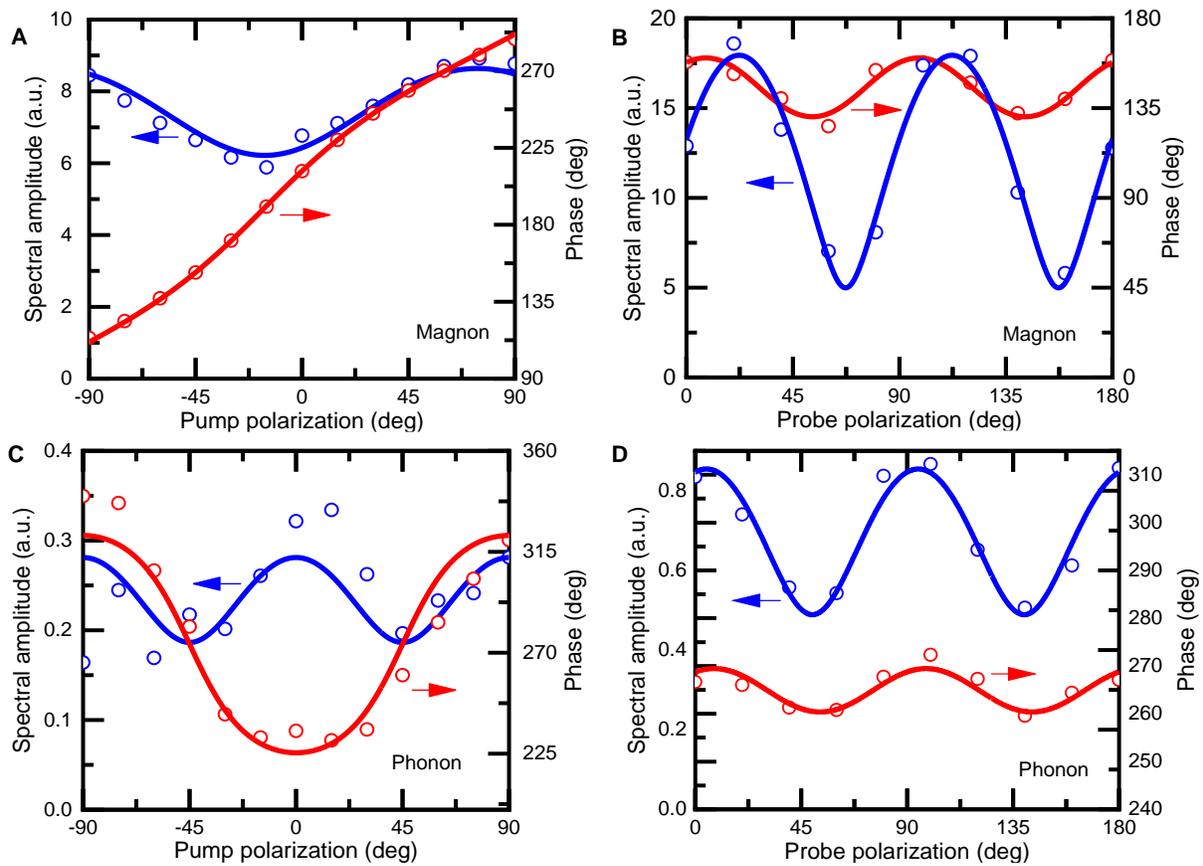

**Fig. S5. Comparison of experimental and theoretical selection rules.** Selection rules analysis of the magnon and phonon modes at $T = 6$ K. The amplitude and the phase of the magnon (**A**) and the phonon (**C**) modes as a function of the THz magnetic field orientation angle $\psi$ at the horizontal probe electric field ($\gamma = 90°$). The amplitude and the phase of the magnon mode (**B**) and the phonon (**D**) modes as a function of the probe electric field orientation angle $\gamma$ at the horizontal THz magnetic field ($\psi = 90°$). The origin data are plotted in fig. S6 and S7. The fits are based on the Eqs. 30-33 and Eq. 44 of Supplemental material.

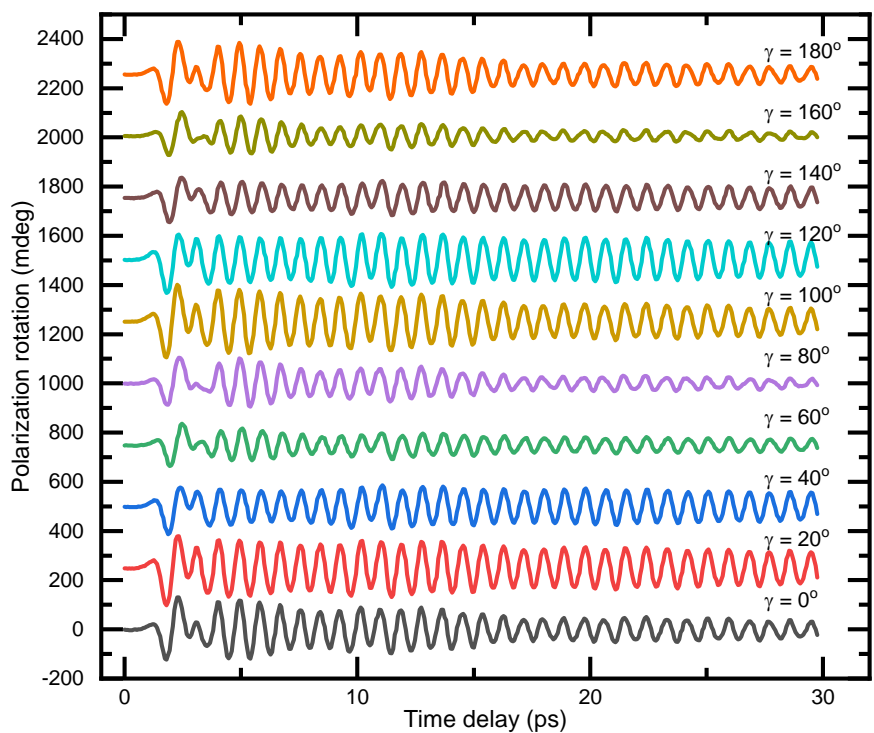

**Fig. S6. THz induced dynamics on the probe electric field orientation.** Terahertz induced probe polarization rotation measured for the different probe electric field orientation angle γ shown near corresponding curves. $T = 6$ K, the horizontal THz magnetic field ($\psi = 90°$).

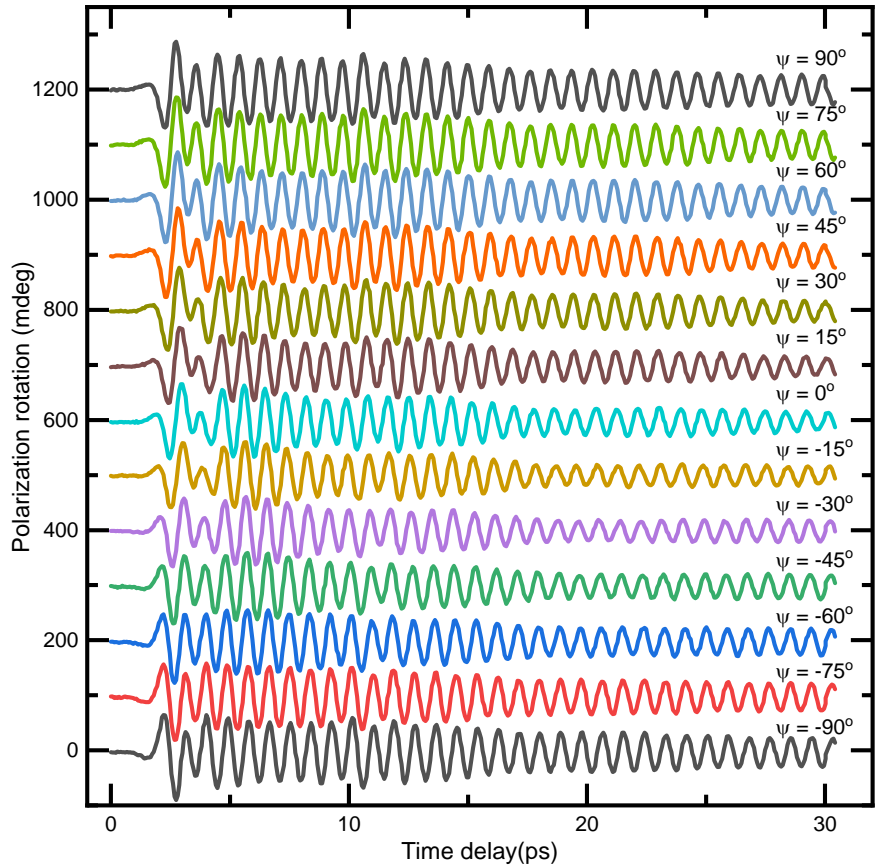

**Fig. S7. THz induced dynamics on the THz pump magnetic field orientation.** Terahertz induced probe polarization rotation measured for the different THz magnetic field orientation angle ψ shown near corresponding curves. $T = 6$ K, the horizontal probe electric field ($\gamma = 90°$).

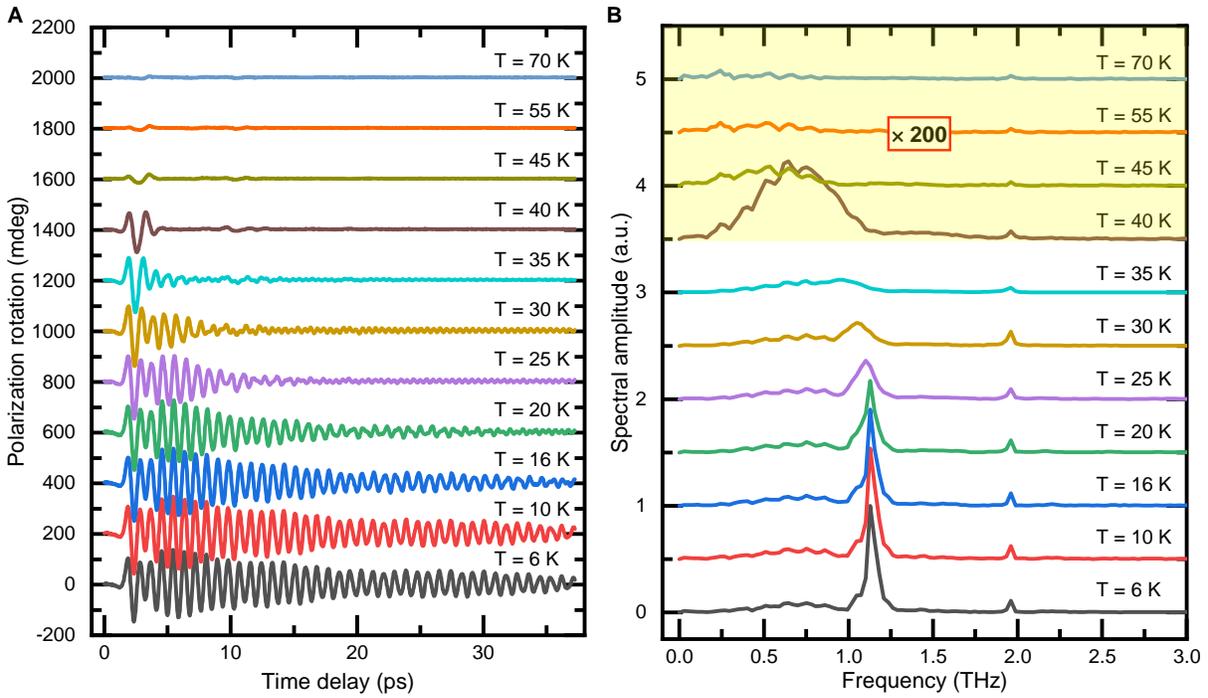

**Fig. S8. THz induced dynamics on temperature.** Terahertz induced probe polarization rotation waveforms (**A**) and corresponding Fourier spectra (**B**) at different temperatures shown near corresponding curves. $T$ = 6 K, the horizontal probe electric field ($\gamma$ = 90°), and the horizontal THz magnetic field ($\psi$ = 90°). Due to different dumpings of the magnon and phonon modes, the phonon becomes clearly visible in time domain at 30 K. Near $T_N$ = 39 K the magnon mode disappears, while the phonon mode experinces significant drop on an amplitude.

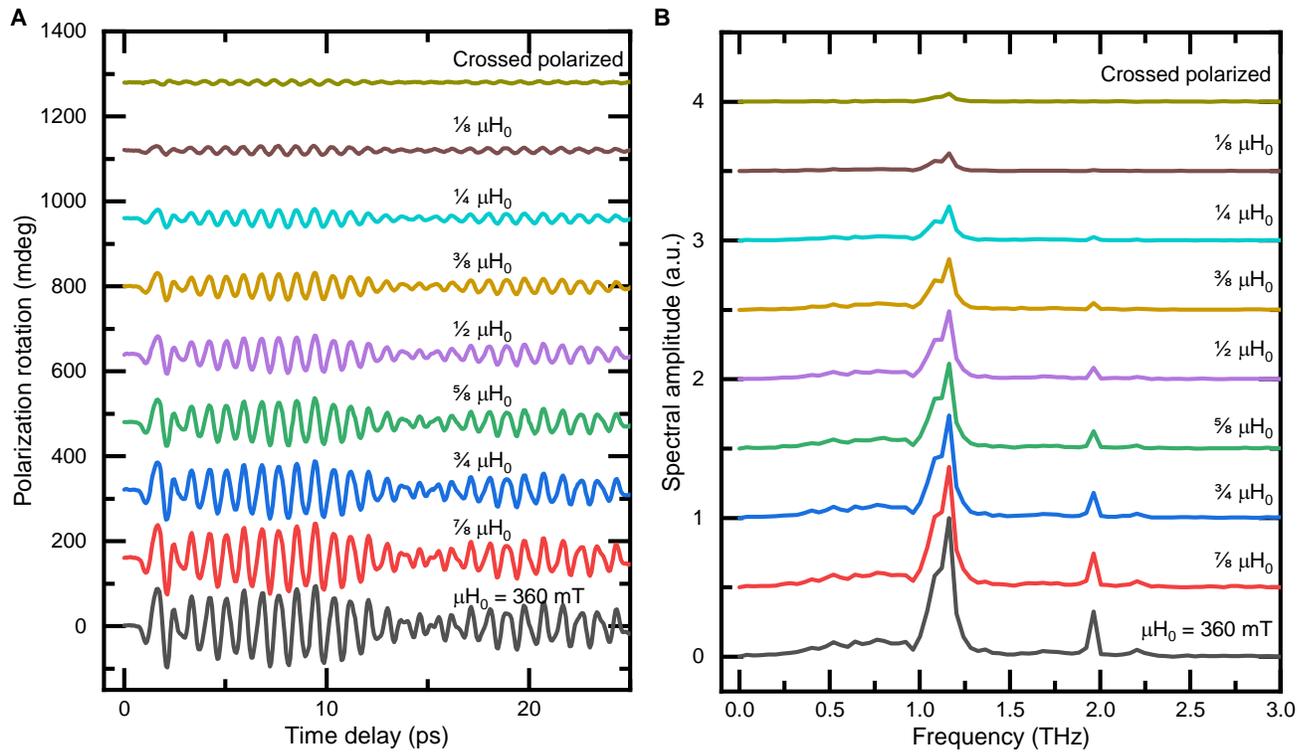

**Fig. S9. THz induced dynamics on the THz field strength.** Terahertz induced probe polarization rotation waveforms (**A**) and corresponding Fourier spectra (**B**) on the different THz pump magnetic field strength $\mu H_0$ shown near corresponding curves. $T = 6$ K; the horizontal probe electric field ($\gamma = 90°$) and the horizontal THz magnetic field ($\psi = 90°$). The remaining signal at crossed polarizations can be result of slight grid-polarizers axes misalignment or initial ellipticity of the THz pulse generated in the LiNbO$_3$ crystal.

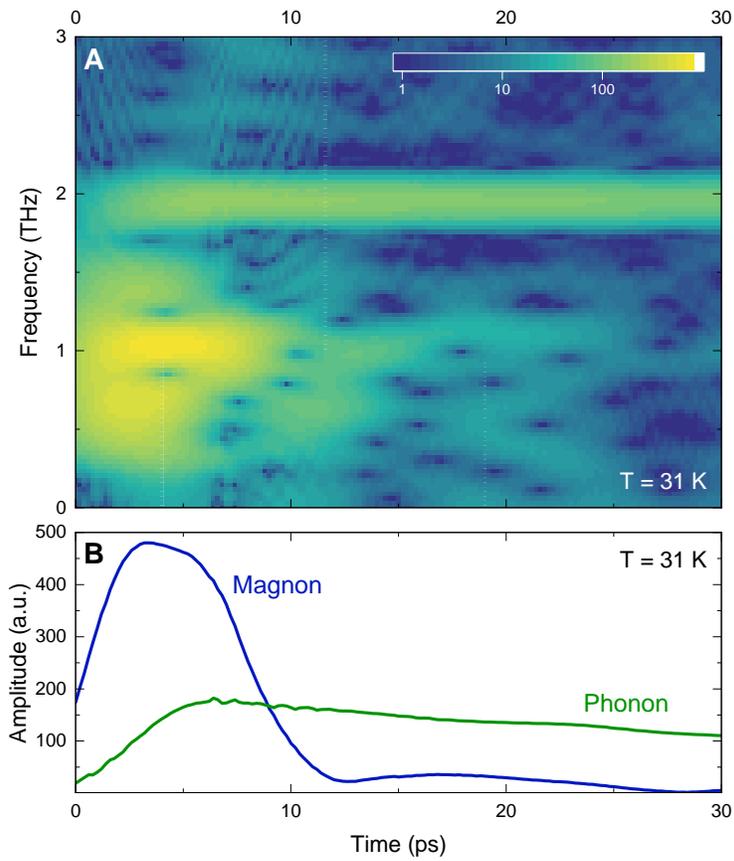

**Fig. S10 Sliding Fourier transform.** (**A**) Spectrogram of the waveform in fig. S8A at 31 K. (**B**) The slices of the spectrogram at $\Omega_m$ (blue curve) and $\Omega_{ph}$ (green curve). Excitation of the $B_{1g}$ phonon is appearing after the magnon excitation. Nonzero amplitudes at zero time are the result of the finite length window in the short-time Fourier transform.

| Operator | $E_x$ | $E_y$ | $H_x$ | $H_y$ | $I_x$ | $I_y$ | $L_{0z}$ | $E_x^2 - E_y^2$ | $I_x H_y - I_y H_x$ | $M_z$ | $\sigma_{xy}$ |
|---|---|---|---|---|---|---|---|---|---|---|---|
| $4\underline{z}$ | $-E_y$ | $E_x$ | $H_y$ | $-H_x$ | $-I_y$ | $I_x$ | $-L_{0z}$ | $-(E_x^2 - E_y^2)$ | $-(I_x H_y - I_y H_x)$ | $M_z$ | $-\sigma_{xy}$ |
| $2d$ | $-E_y$ | $-E_y$ | $H_y$ | $H_x$ | $I_y$ | $I_x$ | $-L_{0z}$ | $-(E_x^2 - E_y^2)$ | $-(I_x H_y - I_y H_x)$ | $-M_z$ | $\sigma_{xy}$ |

**Table S1.** Transformation of different parameters under group operators of 4/mmm point group.